\documentstyle[aps,prl,twocolumn,psfig]{revtex}

\draft
\begin{document}
\newcommand {\be}{\begin{equation}}
\newcommand {\ee}{\end{equation}}
\newcommand {\bea}{\begin{eqnarray}}
\newcommand {\eea}{\end{eqnarray}}
\newcommand {\nn}{\nonumber}

\twocolumn[\hsize\textwidth\columnwidth\hsize\csname@twocolumnfalse%
\endcsname

\title{Order by disorder from non-magnetic impurities in a 
two-dimensional quantum spin liquid}

\author{S. Wessel$^1$, B. Normand$^2$, M. Sigrist$^3$, and 
S. Haas$^1$}
\address{$^1$Department of Physics and Astronomy, University of Southern
California, Los Angeles, CA 90089-0484\\
$^2$ Theoretische Physik III, Universit\"at Augsburg, D-86135 Augsburg,
Germany\\
$^3$ 
Yukawa Institute for Theoretical Physics, Kyoto University, Kyoto 606-8502,
Japan
}

\date{\today}
\maketitle

\begin{abstract}

We consider doping of non-magnetic impurities in the spin-1/2, 
1/5-depleted square lattice. This structure, whose undoped phase 
diagram offers both magnetically ordered and spin-liquid ground 
states, is realized physically in CaV$_4$O$_9$. Doping into the ordered 
phase results in a progressive loss of order, which becomes complete at 
the percolation threshold. By contrast, non-magnetic impurities 
introduced in the spin liquids create a phase of weak but long-ranged 
antiferromagnetic order coexisting with the gapped state. The latter 
may be viewed as a true order-by-disorder phenomenon. We study the 
phase diagram of the doped system by computing the static 
susceptibility and staggered magnetization using a stochastic 
series-expansion quantum Monte Carlo technique. 

\end{abstract}
]

Doping of low-dimensional quantum spin liquids leads to a 
variety of phenomena, including unconventional superconductivity
when localized spins are replaced by mobile holes \cite{dagotto}, 
and effective random spin systems, whose low-temperature properties 
are dominated by large effective spin degrees of freedom, in the 
presence of static impurities\cite{westerberg,sigrist,haas}. Such 
low-temperature phases have been observed recently in 
quasi-one-dimensional (1d) spin systems. In the two-leg ladder 
compound Ca$_{14-x}$Sr$_x$Cu$_{24}$O$_{41}$, a superconducting 
transition has been measured at high pressure \cite{akimitsu}. 
Zn-doping of the quasi-1d ladder material $\rm SrCu_2O_3$ 
\cite{azuma}, and of the dimerized chain CuGeO$_3$ \cite{rskumhhs}, 
yields low-temperature phases with weak antiferromagnetic (AF) 
order, arising from effective couplings of random strength and 
sign between unpaired spins.

The question naturally arises of whether similar phenomena can be 
expected in quasi-2d systems. There are several prominent 
examples of 2d spin liquids, including (i) structurally frustrated 
materials, such as $\rm SrCu_2(BO_3)_2$ \cite{rkysmokksgu}, 
containing AF coupled spin-1/2 triangles, (ii) interaction-frustrated 
systems, such as the $J_1$-$J_2$ Heisenberg model on a square lattice 
\cite{rdm} in the gapped regime around $J_2/J_1 = 0.5$, and (iii) 
structurally discretized materials, such as the 1/5-depleted AF square 
lattice formed by the spin-1/2 vanadium ions of CaV$_4$O$_9$ 
(Fig. 1) \cite{cavo}. These systems have resonant-valence-bond 
(RVB) ground states with a finite spin gap. Their AF correlation 
length is finite, decreasing with increasing gap size, and the 
low-temperature uniform susceptibility is exponentially activated. 
We find that non-magnetic impurities in such systems act to 
induce a long-ranged AF order, which exists on the subsystem of 
uncombined spins, and coexists with the dominant, gapped phase of 
the spin liquid. This order arising from disorder is the central 
result of our analysis. As in the quasi-1d case \cite{rfts}, its origin 
lies in the unfrustrated nature of the exponentially weak interactions 
between unpaired spins introduced when impurities break the RVB singlet 
units in a bipartite lattice: these are ferromagnetic when the free 
spins lie on the same sublattice, and AF for opposite sublattices 
\cite{sigrist}. 

\begin{figure}[h]
\centerline{\psfig{figure=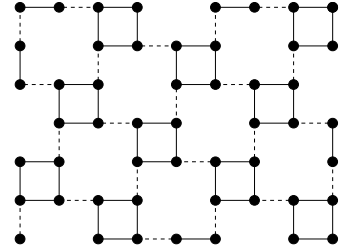,width=4.3cm,angle=0}}
\vspace{0.3cm}
\caption{ 
The 1/5-depleted square lattice. Intra-plaquette couplings $J$ 
are indicated by solid lines, and inter-plaquette couplings 
$J^{\prime}$ by dashed lines. }
\end{figure}

In this work we consider the effects of non-magnetic impurities in 
the 1/5-depleted square lattice geometry (Fig. 1), motivated by the 
structure of CaV$_4$O$_9$, and by the possibility of doping this 
particular material with Ti. The set of superexchange parameters 
required to model CaV$_4$O$_9$ has been found \cite{rkhskktyskmn,rcp} 
to be dominated by the frustrating, next-neighbor AF interaction,
which to a reasonable approximation allows the real system to be 
considered as an ensemble of weakly coupled metaplaquettes (a 
square of side $\sqrt{2}$ larger than the nearest-neighbor spacing) 
of 4 V$^{4+}$ spins. We present first a theoretical analysis for the 
unfrustrated system, which contains all the qualitative features 
we wish to address, and return thereafter to the experimental situation. 

The effective magnetic Hamiltonian, 
\bea
H = J \sum_{\Box } {\bf S}_i\cdot {\bf S}_j + 
J^{\prime} \sum_{\Box - \Box } {\bf S}_i\cdot {\bf S}_j,
\eea
is a spin-1/2 AF Heisenberg model with $J$ is the intra-plaquette and 
$J^{\prime}$ the inter-plaquette exchange coupling constant, in which 
the sums marked by
$\Box$ and $\Box - \Box$ run over nearest neighbors within and between
the plaquettes. We will explore the full parameter range  $J^{\prime}/J 
\ge 0$, which contains a spin-liquid, plaquette RVB (PRVB) phase for 
$J^{\prime}/J \alt 0.93$ , an AF-ordered phase in the approximate
interval $J^{\prime}/J \in [0.93,1.68]$, and a further spin liquid, 
the dimer RVB (DRVB) phase, at larger $J^{\prime}/J \agt 1.68$ 
\cite{troyer3}. Expansions about the plaquette and dimer limits 
will be compared with simulations using
the stochastic series-expansion quantum Monte Carlo (SSE QMC) algorithm 
\cite{sandvik1}, on 1/5-depleted, square lattices of up to 30$\times$30 
sites, and at temperatures down to T = 0.02$J$. This algorithm 
involves expansion of the partition function in inverse temperature, 
employs local and global system updates, and is significantly more 
efficient than conventional QMC schemes. In the presence of non-magnetic 
impurities, ensemble averages over typically 500 random realizations were 
taken to ensure numerical errors smaller than the symbol sizes in Figs. 2 
and 3. 
 
\begin{figure}[h]
\centerline{\psfig{figure=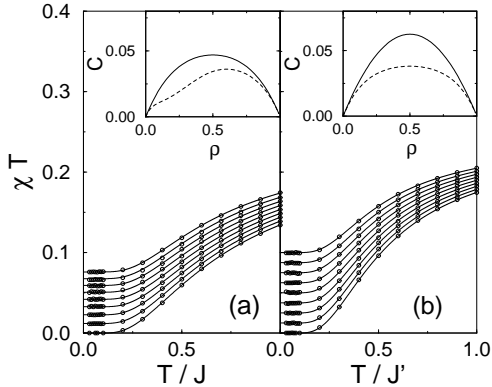,width=6cm,angle=270}}
\vspace{0.5cm}
\caption{
(a) Ensemble of decoupled plaquettes: temperature dependence of
$T\chi(T)$ at impurity concentrations $\rho = 0.0, 0.05, ..., 0.4$. 
Solid lines are exact results, and symbols are QMC simulations. 
$T\chi(T)$ is normalized by the number of spins in the system.
(b) As (a), for decoupled dimers. Insets: dependence of 
zero-temperature Curie constant $C$ on $\rho$. Solid lines denote 
decoupled plaquette and dimer limits, and dashed lines the results 
from perturbation expansion with (a) $J^{\prime}/J = 0.1$ and (b) 
$J/J^{\prime} = 0.1$. }
\end{figure}

Considering first an ensemble of completely decoupled 4-spin plaquettes, 
a small number of randomly placed, non-magnetic dopants creates free 
spin-1/2 moments, corresponding to the doubly degenerate $S^z=\pm 1/2$
ground state of 3-spin Heisenberg clusters \cite{laukamp}. Each of 
these ``pruned'' plaquettes gives a paramagnetic contribution to the 
susceptibility $\chi(T)$ at low temperature. Upon further doping, 
clusters containing $n < 3$ spins are also created, with distributions 
$P(n)$ \cite{fn}, but only unpaired spins contribute to the low-$T$ Curie 
tail. Fig. 2(a) shows $T\chi(T)$ for various impurity concentrations. 
Without impurities, this quantity extrapolates to zero as the temperature is 
lowered, indicative of the spin gap in the clean system. With impurities, 
the zero-temperature Curie constant is given by  
\bea
C = \lim_{T\rightarrow 0} [T\chi(T)] = \frac{1}{4} \left( \frac{P(1) 
 + 1/3 P(2) + P(3)} {1 - \rho}\right),
\eea 
and depends on concentration as shown in the inset of Fig. 2(a).
For all $\rho$, three distinct temperature regimes are observed.
(i) For $T \gg \Delta_{\rm max}$ (not shown) the spins are essentially 
independent, each contributing $\frac{\mu_B^2}{4k_B}$ to $C$. 
$\Delta_{\rm max} = 1.5 J$ is the largest energy level separation in 
any cluster. 
(ii) For $\Delta_{\rm min} < T < \Delta_{\rm max}$, some spins freeze 
into singlets, gradually reducing the overall magnetic response by a 
factor of approximately $\rho$ as $T$ decreases. $\Delta_{\rm min} = 
0.25J $ is the smallest energy spacing in any cluster. 
(iii) For $T < \Delta_{\rm min}$, plateaus appear in $T\chi(T)$ when  
the contributing clusters reach their ground states. 

\begin{figure}[h]
\centerline{\psfig{figure=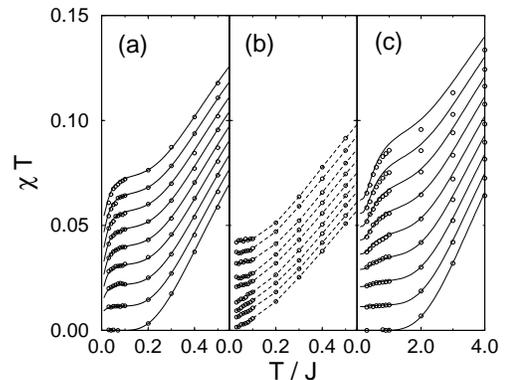,width=6cm,angle=270}}
\vspace{0.3cm}
\caption{ Temperature dependence of $T\chi(T)$ at impurity concentrations 
$\rho = 0.0, 0.05, \dots, 0.4$. (a) PRVB regime, $J^{\prime}/J = 
0.3$; (b) ordered AF regime, $J^{\prime}/J = 1.2$; (c) DRVB regime, 
$J^{\prime}/J = 2.0$. Solid lines are results from expansions 
about the decoupled limits, and symbols are from SSE QMC simulations. 
Dashed lines in (b) are guides to the eye. }
\end{figure}

With finite inter-plaquette coupling, spin-1/2 moments on adjacent 
plaquettes may combine to form new RVB clusters with smaller spin 
gaps, involving larger numbers of spins. In the susceptibility, 
this effective recombination reduces $C$ from its value for decoupled, 
pruned plaquettes, further suppressing $T \chi(T)$ at very low $T$. 
This cluster formation may be quantified from the distribution 
functions for neighboring plaquettes, evaluated perturbatively in 
$J^{\prime}/J$. These functions have a non-trivial dependence on the 
relative positions of the dopant vacancies in adjacent clusters.
Close to the PRVB limit, the inset of Fig. 2(a) illustrates this 
reduction of $C$, which is strongest at smaller $\rho$ where more spins 
are available to form larger RVB clusters. Fig. 2(b) shows analogous 
results for the DRVB limit. 

In Fig. 3, the $T$-dependence of the uniform susceptibility 
is shown at various doping levels in the three regimes of
$J^{\prime}/J$. In Figs. 3(a) and (c), results from SSE QMC 
simulations (symbols) are plotted along with those from perturbative 
expansions about the limits of completely decoupled plaquettes and 
dimers (solid lines). At $\rho = 0$, $T \chi(T)$ clearly displays 
activated behavior in the two gapped regions, but a linear form 
in the long-range-ordered phase. 

{\it PRVB:} At small $\rho$, a plateau develops in $T \chi(T)$ 
at low temperatures, $T \sim \Delta_{\rm min}$, indicating a
freezing of the plaquettes into their ground state configurations. 
For a finite inter-plaquette coupling, the additional reduction of 
$C$ observed at ultra-low temperatures indicates that the plaquette 
spins become correlated on an ultra-low energy scale $J_{\rm eff}$. 
At sufficiently small $\rho$, this energy can be estimated \cite{sigrist} 
by assuming RKKY-like interactions between the impurity spins, yielding
\bea
J_{\rm eff} \approx J \exp{\left(-\frac{(1-\rho)}{\xi_{AF}(\rho, T)}\right)},  
\eea
where $\xi_{AF}$ is the short-range AF correlation length in the 
spin-liquid regimes. Approaching $T = 0$, the Curie constant extrapolates 
to $C = \rho/12$, a reduction of roughly 1/3 compared to the plateau value 
\cite{sigrist}. At larger impurity concentrations, the onset of this 
ultra-low-$T$ regime moves up, consistent with $T \sim J_{\rm eff}$ (Eq. 
3). At still higher doping, the picture of isolated impurity spins, and 
RKKY interactions mediated via the spin-liquid RVB background, breaks down. 
The regimes of moment formation and correlation overlap, and the plateau 
feature in $T \chi(T)$ becomes less pronounced. On doping towards the 
percolation threshold $\rho_c$, the magnetic response is dominated by 
the largest percolation clusters, typically containing several pruned 
plaquettes. From simulations in the classical limit \cite{stauffer}, 
we find $\rho_c = 0.26(9)$ for the 1/5-depleted lattice. Studies of 
percolation thresholds in regular square lattices \cite{stauffer} 
suggest that the value of $\rho_c$ for the quantum spin system differs 
by less than $1\%$ from the classical value ($\rho_c = 0.407$). In this 
regime the plaquette plateaus disappear, a trend accentuated as the 
long-range-ordered regime is approached ($J^{\prime}/J \rightarrow 0.93$), 
where $\xi_{AF}$ increases and (Eq. 3) the onset of the ultra-low-$T$ 
regime moves to lower energies. 

{\it DRVB:} As in the PRVB regime, non-magnetic impurities 
introduce local moments in the gapped, spin-liquid state. A dimer 
``plateau'' is reached at $T \sim J_{\rm eff}$, and further moment 
recombination occurs at ultra-low temperatures, leading to an additional 
reduction of $T\chi(T)$ as $T \rightarrow 0$. The plateaus disappear on 
approaching the transition to the AF phase, $J^{\prime}/J \rightarrow 
1.68^+$.

{\it AF:} In contrast to gapped spin liquids, isolated, non-magnetic 
impurities do not introduce quasi-free local moments, because all 
spins remain aligned with the staggered AF background \cite{sandvik}.
However, the low-temperature susceptibility does have a divergent 
contribution, albeit with $C$ rather smaller than in the spin liquids 
at given $\rho$, arising from free moments isolated from the AF system 
by dopants. 
 
\begin{figure}[h]
\centerline{\psfig{figure=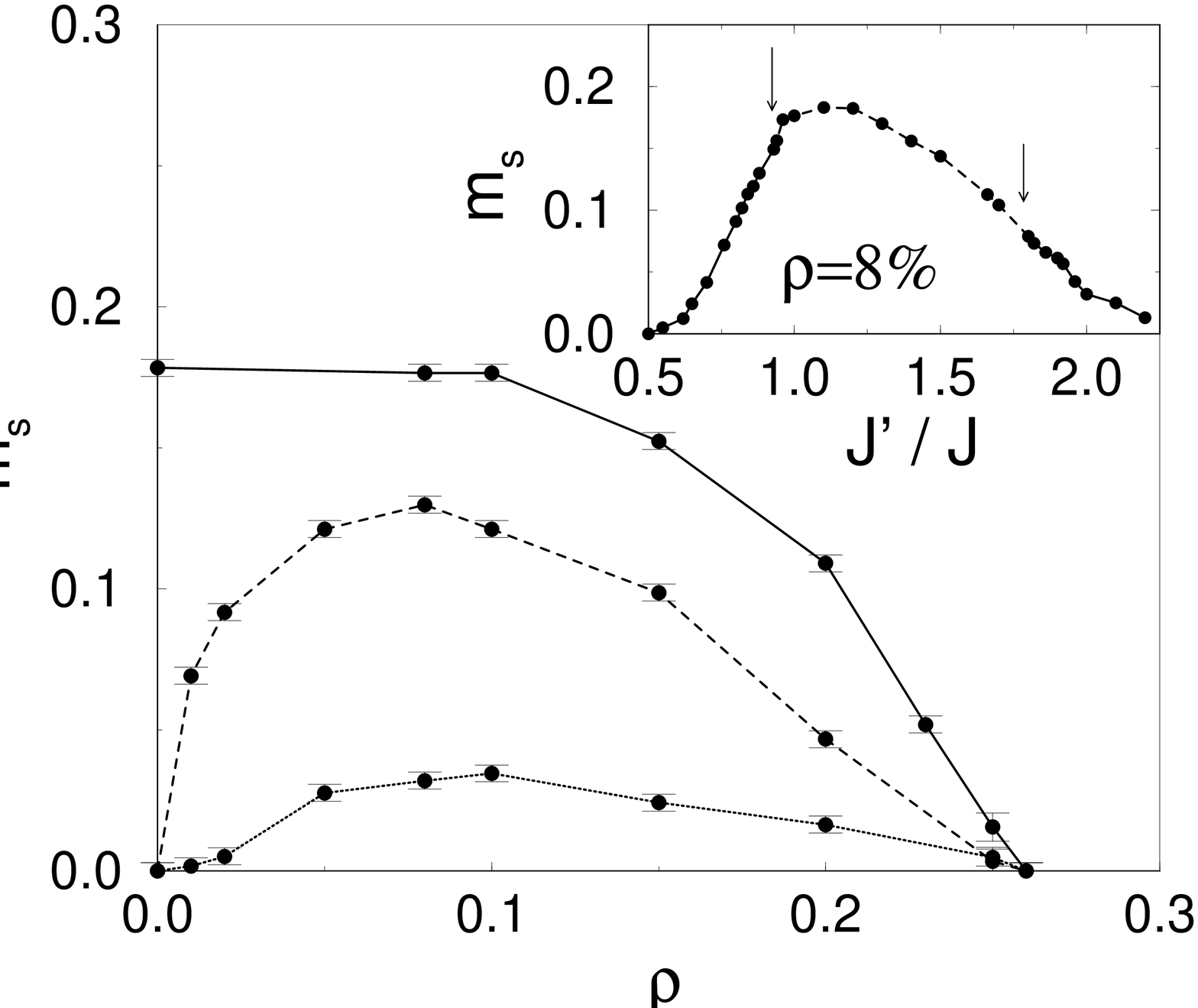,width=6cm,angle=270}}
%\vspace{0.3cm}
\caption{Staggered magnetization as a function of impurity doping 
for coupling ratios $J^{\prime}/J$ in the PRVB (dashed), ordered 
(solid) and DRVB (dotted) regimes. The inset shows the evolution of 
$m_s$ with $J^{\prime}/J$ at a fixed doping $\rho$ = 8\%; arrows 
indicate the phase boundaries of the pure system.}
\end{figure}

Returning to the issue of order by disorder, our simulations close to 
the AF regime of the pure system show that introduced moments do not 
simply recombine into larger RVB clusters, but form a long-range-ordered 
AF network at ultra-low temperatures. This doping-induced long-range 
order is best characterized by the staggered magnetization $m_s$, which 
is shown in Fig. 4 as a function of non-magnetic impurity doping within 
each of the distinct phases of the system. As also in Fig. 5, these data 
were obtained by finite-size extrapolation at ultra-low temperatures, 
and thus represent the thermodynamic limit. In the long-range-ordered 
AF regime, we see that random static vacancies simply reduce the average 
staggered magnetization, driving it to zero beyond the percolation 
threshold $\rho_c = 0.26(9)$. By contrast, in the spin-liquid regimes 
the doping-induced moments have effective interactions mediated by the 
RVB background, which are {\it unfrustrated} \cite{sigrist}. A long-ranged 
order may then be expected to emerge from the large-spin clusters at 
zero temperature. Indeed, the resulting AF network has 
an extensive staggered magnetization, which peaks around $\rho = 0.08$. 
This ordered AF phase coexists \cite{rfts} with the gapped, spin-liquid 
state of the majority, undoped plaquettes or dimers (Fig. 3). The mechanism 
leading to this moment formation may be viewed as an order-by-disorder 
phenomenon. The inset of Fig. 4 shows the staggered magnetization as a 
function of the coupling ratio at the maximal doping. The disorder-induced, 
ordered moments in the spin-liquid regions are significant over a wide 
range of $J^{\prime}/J$, and become large near the phase boundaries of 
the pure system, where they cross continuously to the full moment of the 
intrinsically ordered regime. 

Our results are summarized by the phase diagram in Fig. 5, where the 
shading illustrates the strength of the staggered magnetic order. The 
solid lines depicting phase boundaries may not be taken as true 
transitions at any finite doping: impurity-induced AF order within a 
spin liquid crosses smoothly to impurity-suppressed order within an AF, 
with no vanishing order parameter. Impurities may be considered as 
damaging both to the AF ordered phase, and the quantum spin liquid phase. 

\vspace{-0.5cm}
\begin{figure}[h]
\centerline{\psfig{figure=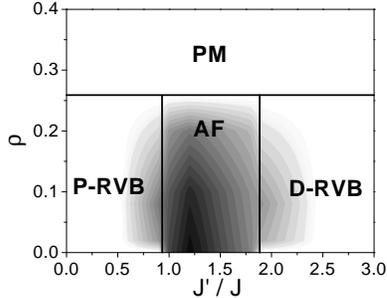,width=6.2cm,angle=0}}
%\vspace{0.3cm}
\caption{Phase diagram for the 1/5-depleted, square-lattice 
antiferromagnet with doping by non-magnetic impurities. }
\end{figure}

Experimentally, the situation in CaV$_4$O$_9$ may be more complex 
than that simulated here. Inelastic neutron scattering measurements 
\cite{rkhskktyskmn}, and subsequent refinements by comparison with 
simulations \cite{rcp}, show that the magnetic system is composed of 
metaplaquettes with $J_2 = J$ = 14meV, which interact with their neighbors 
via mutually frustrating couplings $J_1 = J_1^{\prime}$ = 0.49$J$, 
and with next neighbors by a further weak coupling $J_2^{\prime} = 
0.25 J$. Because of the frustrated nature of the subdominant 
interactions $J_1$, it would appear reasonable that the effective coupling 
$J^{\prime}$ between metaplaquettes is small, and thus that the 
physical system is well in the PRVB regime; this deduction is 
consistent with the robust spin gap, $\Delta = 9.4$meV, of CaV$_4$O$_9$. 
Doping with non-magnetic impurities could be effected by random 
replacement of some spin-1/2 V$^{4+}$ ions by non-magnetic Ti$^{4+}$. 
Our simulations of the uniform susceptibility and staggered 
magnetization suggest that the weak AF order induced on 
doping should be detectable in 5-10\% Ti-doped CaV$_4$O$_9$, by 
measurements of $\chi(T)$, Knight shift or magnetic Bragg scattering. 

In summary, we have analyzed the behavior on doping by non-magnetic 
impurities of the spin-liquid and ordered AF phases of the Heisenberg 
model on a 1/5-depleted, square lattice of spins S = 1/2. In the spin 
liquids, introduction of impurities creates effectively free spins which 
combine through exponentially weak interactions to create a long-ranged AF 
order at $T = 0$, coexisting with the gapped, spin-liquid phase. In the 
intrinsically AF regime, impurities cause a progressive destruction of 
the long-range order until it vanishes at the percolation threshold.  
The SSE QMC algorithm which we employ proves to be a powerful and 
efficient means of simulating the properties of a spin system; these 
attributes are particularly important when many realizations of random 
impurity distributions are required. 
Finally, the unique magnetic plaquette structure of CaV$_4$O$_9$ 
provides a well-defined starting point for studying the effect of
non-magnetic impurities in spin liquids, and provides valuable 
insight for other 2d RVB compounds, such as frustrated Heisenberg 
systems, where a recombination of pruned bonds into new RVB clusters 
is also expected to occur.

We acknowledge the financial support of the Zumberge Foundation (SW, SH), 
SFB 484 of the Deutsche Forschungsgemeinschaft (BN), and the Japanese
Ministry of Education and Science (MS).

\end{document}